\documentclass[11pt,letterpaper]{article}

\usepackage{graphicx}
\usepackage{bm}
\usepackage{verbatim}
\usepackage{subfigure}
\usepackage{epsfig}
\usepackage{cite}
\usepackage{rotating}
\usepackage{balance}
\usepackage{authblk}

\begin{document}

\title{PowerAI DDL}
\author{Minsik Cho, Ulrich Finkler, Sameer Kumar, David Kung, Vaibhav Saxena, Dheeraj Sreedhar}
\affil{IBM Research}

\maketitle

\begin{abstract}
    As deep neural networks become more complex and input data-sets grow larger, it can take days or
even weeks to train a deep neural network to the desired accuracy. Therefore, distributed Deep Learning at a
massive scale is a critical capability, since it offers the potential to reduce the training
time from weeks to hours. In this paper, we present a software-hardware co-optimized distributed
Deep Learning system that can achieve near-linear scaling up to hundreds of GPUs. The
core algorithm is a multi-ring communication pattern that provides a good tradeoff between
latency and bandwidth and adapts to a variety of system configurations. The communication
algorithm is implemented as a library for easy use. This library has been integrated into Tensorflow, Caffe, and Torch.
We train Resnet-101 on Imagenet 22K with 64 IBM Power8 S822LC servers (256 GPUs) in about 7 hours to an accuracy of 33.8\%
validation accuracy. Microsoft's ADAM \cite{ADAM} and Google's DistBelief \cite{DistBelief} results did not reach
30\% validation accuracy for Imagenet 22K. Compared to
Facebook's recent paper\cite{FACEBOOK-256} on 256 GPU training, we use a different communication
algorithm, and our combined software and hardware system offers better communication overhead for Resnet-50.
A PowerAI DDL enabled version of Torch completed 90 epochs of training on Resnet 50 for 1K classes 
in 50 minutes using 64 IBM Power8 S822LC servers (256 GPUs).

\end{abstract}


\section{Introduction}

Deep Learning has become the de facto technique for an increasing number of cognitive applications, including vision,
speech and language translation. The success is driven by the availability of an enormous volume of data and advances
in deep neural networks, but it is these enabling factors that make Deep Learning one of the most computationally-demanding
HPC applications. Hardware accelerators such as GPUs and their accompanying software stack have provided a significant amount
of speed up. However, deep neural network training for speech and vision can still take days and even weeks. Therefore
parallelization by distributing the Deep Learning training to many (upwards of hundreds) GPUs over a compute cluster is
critical to cut the training time from weeks to hours.

Distributed Deep Learning (DDL) is challenging because as the number of learners increases, 
the computation time decreases whereas
the amount of communication stays constant, resulting in unfavorable computation to communication ratios. Successful distributed
Deep Learning requires an infrastructure in which the hardware and software are co-optimized to balance the computational
requirements with the communication demand and interconnect bandwidth. In addition, the communication latency plays an important
role in massive scaling of GPUs (over 100). If these factors are not kept under control, distributed Deep Learning can quickly
reach the point of diminishing return.

In this paper, we present a software-hardware co-optimized distributed Deep Learning system that can achieve near-linear scaling up to 256 GPUs. The communication algorithm is encapsulated as a communication library, called PowerAI DDL, that provides a set of communication APIs for implementing Deep Learning communication patterns as an integral part of any Deep Learning framework. Currently, PowerAI DDL has been integrated with 
{\em Tensorflow}\cite{TF}, {\em Caffe}\cite{CAFFE}, and {\em Torch}\cite{TORCH}. 
PowerAI DDL provides a unified infrastructure for distributed Deep Learning over multiple GPUs for a single node, multiple nodes and a cloud environment. PowerAI DDL leverages an innovative multi-ring communication algorithm that balances communication latency with the communication overhead. It adapts to the hierarchy of communication bandwidths in any systems -- bandwidths that are often highest within a node and then subsequently decreases between nodes and then between racks in a cluster. PowerAI DDL's adaptation enables the ``optimal" distributed Deep Learning solution.

The paper is organized as follows. In section \ref{sec-prior}, we review prior art, focusing on the distributed Deep Learning
result that Facebook AI Research \cite{FACEBOOK-256} recently announced. In section \ref{sec-method}, 
we discuss what the methodology for benchmarking distributed Deep Learning
should be. In section \ref{sec-alg}, we describe PowerAI DDL and our multi-ring algorithm. In section \ref{sec-exp},
we present and analyze our
results on distributed Deep Learning using 64 IBM Power8 S822LC servers (256 GPUs).

\section{Prior Art} \label{sec-prior}

Massive scale distributed Deep Learning has been the holy grail of the deep learning research community for the past several years. There have been many attempts and reports on scaling to 64, 128 GPUs, but most of them only address the problem from a specific angle. In our opinion, the recent paper by Facebook\cite{FACEBOOK-256} is the first attack on 256 GPUs that addresses the entire massive scale Deep Learning problem, and provides truly impressive practical results. One thing we want to highlight is that the success of Facebook's paper hinges on being able to train Resnet-50 to the best accuracy while using a very large batch size, which runs counter to conventional wisdom. To train a neural network using 256 GPUs, one must solve the problem of converging to good accuracy with large batch sizes. The GPU computational efficiency at small batch size (4 or less per GPU) is very sub-par, and the compute to communication ratio is also too low for scaling to be efficient. Therefore the batch size per iteration must be in the thousands. The Facebook paper proffered a linear scaling conjecture for the learning rate for large batch size and selected a batch size of 8192 for the experiment, and demonstrated that it works, at least for Resnet-50. The linear scaling conjecture is plausible if the operating region in the parameter space is rather linear, which is not unreasonable when the training process has been warmed-up, as is being done in the paper. This paper\cite{FACEBOOK-256} is the first instance of a successful training of an "industrial" strength neural network to the best published accuracy with such a large batch size.

Other relevant details of \cite{FACEBOOK-256} are as follows. The hardware platform is 32 DGX-1 systems \cite{DGX1} (256 P100-SXM2) and the communication algorithm used is the recursive halving and recursive doubling algorithm well known in the HPC community. Resnet-50\cite{RESNET} on data set Imagenet 1K is trained using Caffe2 with the usual data augmentation techniques. The learning rate schedule starts with a warm-up phase, followed by a ramp up and then the usual step-wise descent. The scaling efficiency in \cite{FACEBOOK-256} is defined as the ratio between the run time per iteration of the one node case and the run time per iteration for the 256 GPU case, which we believe artificially inflated the scalability metric slightly (as opposed to measuring
scaling based on 1 GPU). The validation accuracy in \cite{FACEBOOK-256} reaches ~76\% in a hour, with scaling efficiency at ~90\%.

\section{ Distributed Deep Learning Benchmarking Methodology} \label{sec-method}

To compare the efficacy of different distributed Deep Learning systems, we need to capture the various characteristics of the distributed Deep Learning problem. Several obvious candidates come to mind, namely, scaling efficiency, accuracy, end to end training time, the neural network itself, the GPU being used, and the Deep Learning framework used to perform the training. We will discuss these characteristics separately.

\subsection{Scaling efficiency}

Given a specific batch size and distribution over n GPUs, scaling efficiency is defined as the ratio between the run time of one iteration on a single GPU and the run time of one iteration when distributed over n GPUs. This measurement is often extracted without any training to a given accuracy. Sometimes it is even calculated using synthetic data sets. In such cases, scaling efficiency on its own is completely meaningless, because one can satisfy any given scaling efficiency for any neural network by increasing the batch size. Using too big a batch size will result in converging to an unacceptable accuracy or no convergence at all, or such a long training time as to defeat the purpose of speeding up the training using distribution. Therefore, a high scaling efficiency without being backed up by convergence to a good accuracy and end to end training time is vacuous. In this paper, we only present results for neural network runs with demonstrated successful convergence -- no artificial data or pure per-iteration training comparisons are used.

\subsection{Accuracy and end to end training time}

In comparing two different distributed Deep Learning systems, it is imperative to have both system train to the same accuracy. A few percent improvement in accuracy can convert to adding many days of training time. Ultimately, users are interested in reducing the end to end training time at the same accuracy. Furthermore, accuracy has to be determined with a test data set of sufficient size that contains {\em all} classes used in network training and contains only images that have not been using during network training. Quoting training accuracy is unacceptable, since that just reflects over-fitting and over 90\% training accuracy can be readily achieved for many neural networks. We have observed several unpublished (blog) claims of high neural network accuracy, which does not comply with this methodology. Results in this paper follow this methodology, as did \cite{FACEBOOK-256}.

\subsection{The neural network}

The compute to communication ratio can vary widely for different neural networks. Using a neural network with high compute to communication ratio can hide the ills of an inferior distributed Deep Learning system. If there is not much to communicate, then having a sub-optimal communication algorithm or low bandwidth interconnect will not matter that much. Therefore it is important to select popular neural networks that are widely used for practical applications when performing the comparison.

\subsection{The Deep Learning framework}

Currently, the computation time for one Deep Learning iteration can vary by up to 50\% when different Deep Learning frameworks are being used. This increases the compute to communication ratio and gives the inferior distributed Deep Learning system an unfair uplift to the scaling efficiency. Therefore it is important to use a computationally-efficient Deep Learning framework to compare the merits of different Deep Learning systems. We have integrated PowerAI DDL into multiple frameworks, to demonstrate its robustness and generality.

\subsection{The GPU}

Nvidia P100 GPUs are approximately 3X faster than Nvidia K40 GPUs. Using a slower GPU increases the compute to communication ratio and again gives the inferior distributed Deep Learning system an unfair uplift to the scaling efficiency. When evaluating the communication algorithm and the interconnect capability of a Deep Learning system, it is important to use a high performance GPU.

\subsection{The communication overhead}

The communication overhead is the run time of one iteration when distributed over n GPUs minus  the run time of one iteration on a single GPU. This includes the communication latency and the time it takes to send the message (gradients) among  the GPUs. However it is not equal to the sum of the above quantities since the communication and latency could be hidden using clever algorithms. The communication overhead gives an indication of the quality of the communication algorithm and the interconnect bandwidth.
To perform a meaningful comparison between distributed Deep Learning systems, it is important to start with a popular neural network that has been widely trained, use a Deep Learning framework that is computationally-efficient, and train to best accuracy on high performance GPUs. The measurements to be reported should include the accuracy achieved, the training time it takes to attain that accuracy, the scaling efficiency, and the communication overhead. The overall metric will depend on whether factors such as power efficiency, cost and so on. An example of a meaningful comparison is to use the same number of the same GPUs, same Deep Learning framework, same neural network and the exact same training script and strategy, and compare the run time it takes to train to the same best published accuracy. We follow this methodology in this work and show our contributions to be robust at these standards.

\section{The PowerAI DDL Library} \label{sec-alg}

The PowerAI DDL library provides functionality for high-performance distributed Deep Learning
that can be employed in multiple frameworks, currently there are PowerAI DDL enabled versions of {\em Caffe},
{\em Tensorflow} and {\em Torch}.

\subsection{PowerAI DDL objects}

PowerAI DDL is based around the concept of {\em PowerAI DDL Objects}(). An instance of a PowerAI DDL object
combines a data structure, for example a tensor of gradients, with meta-data. The meta-data define the host and type of memory
the data structure resides in as well as the type of device, e.g. a GPU, and device identifier on that host.

PowerAI DDL objects are analogous to variables and all PowerAI DDL functionality operates on these objects.
The most basic operation is an assignment, where object $A$ is assigned the value contained in object $B$,
creating a copy. The distributed Deep Learning applications employ the operations {\em ReduceScatter} and
{\em AllGather} that operate on a set of variables that are distributed over a set of GPUs in a set of machines.
For example, gradient reduction across 256 GPUs employs 256 PowerAI DDL objects distributed over
64 Power8 S822LC servers, one per process in the application.

\subsection{Current implementation}

The current PowerAI DDL implementation is based on MPI\cite{MPI}, since MPI provides many needed facilities, from scheduling
processes to basic communication primitives, in a portable, efficient and mature software ecosystem.
We select Spectrum MPI, which offers functionality optimized for IBM's Power systems and Infiniband networks, leveraging the
considerable investment in the Spectrum products. The core API can be implemented without MPI if desired.

\subsection{Topology aware communication}

 To synchronize the gradients $\vec{V_j}, j \in \{1\ldots N\}$ of $N$ learners which are distributed
 over $N$ GPUs, a reduction
 \[
 \vec{R} = \sum_{j=1}^N \vec{V_j}
 \]
 has to be performed and the  result $\vec{R}$ has to be distributed to all the learners. For a complex
 neural network such as {\em Resnet-101}\cite{RESNET} for 22000 classes this vector has nearly  90
 million floating point entries, resulting in about 350 Megabyte of data.

\begin{figure}[ht]
\begin{center}
\includegraphics*[width=2.4in]{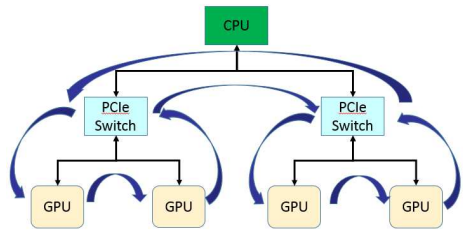}
\end{center}
\caption{Communication ring implemented on a tree topology that uses all links in both directions concurrently. }
\label{fig-treering}
\end{figure}

Deep Learning frameworks often use a tree communication pattern or a parameter server
to perform the reduction among the GPUs. The communication overhead of these techniques has a
logarithmic (or worse) dependence on the number of GPUs and leaves part of the network links 
idle during the reduction process.
On the other hand, the ring (or bucket) reduction\cite{RING}
across $N$ ring nodes requires $N-1$ phases of operation, but it uses all links
concurrently throughout the entire reduction process.
Each phase consists of $N$ concurrent communications, each of which transmits
 $1/N$ of the data set. Hence, the elapsed time of the ring reduction may be estimated by
\[
 T \approx \sum_{j=1}^{N-1} \left ( \frac{S}{N B_{min}}+L \right )
 \]
with $S$ denoting the size of the data set, $B_{min}$ the lowest bandwidth in the ring, and
 $L$ represents the latency incurred to initialize the communications and to synchronize between
 phases.
Figure \ref{fig-treering} shows the basic communication pattern used by PowerAI DDL for a small number of
GPUs, namely, a ring (shown with blue arrows)
in which all nodes communicate concurrently with one other node. Even though the network topology forms a tree,
it can be mapped into a ring topology by using every link in the tree simultaneously in both directions.
The order of the nodes is important in this mapping. As an example, by switching the first node from
the left with the third node from the left causes some network links to be used twice in the same direction,
and thereby increasing the elapsed time. The ring reduction algorithm has a couple of drawbacks - the elapsed time
for the ring reduction algorithm is proportional to the number of GPUs through the latency term, and it is also
dominated by the slowest network link.
While it is a good algorithm for a uniform network when $L$ is small and/or the number of GPUs is small, it is not
the best algorithm when the latency $L$ is large, or the number of GPUs is large, or some network
links are much slower than others.

 Across 256 GPUs, nearly 90 Gigabyte of data have to be transmitted to perform the reduction and
 the same amount of data has to be transmitted to copy the result to all GPUs.
 Even with a fast network connection allowing a net transfer rate of 10 Gigabyte per second, bringing 90 Gigabyte
 of data into a single parameter server would take 9 seconds, not considering the time for the distribution of
 the result. PowerAI DDL achieves the entire reduction and distribution of 350 Megabyte in less than 100 ms by
 using a communication scheme that is optimized with regard to the topology of the network.

When the number of GPUs is large and the network links are not equally fast, PowerAI DDL uses
a multi-dimensional ring algorithm. In the basic ring algorithm, we organize the $N$ GPUs in
a one dimensional structure, a ring. In an $m$-dimensional ring algorithm, we organize the
$N$ GPUs in a $m$-dimensional box, such that $N=d_1*d_2*...*d_m$, where $d_i$ is the size of the
$i^{th}$ dimension of the box. 

A key differentiation of PowerAI DDL is its ability to optimize the communication scheme
based on the bandwidth of each network link, the network topology, and the latency
 for each phase.
For hardware systems with different link speeds, PowerAI DDL adjusts its
dimensions to take advantage of the fastest link, and to prevent the slowest
link to dominate the elapsed time. For example in phase $1$, communication
 is performed via NVLink or memory bus, resulting in a bandwidth of 20 Gbyte/s or more.
 In the second dimension communication may be restricted to a single rack and achieve
 an effective bandwidth of roughly 10 Gbyte/s. The third dimension has to communicate
 between racks and typically achieves a slightly lower bandwidth, for example 9.5 Gbyte/s.
 In a complete ring, the slowest link limits bandwidth to no more than 9.5 Gbyte/s.

\section{Experimental Results} \label{sec-exp}

To evaluate the performance of PowerAI DDL, we used a cluster of 64 IBM Power8 S822LC servers,
each equipped with 4 NVidia Tesla P100-SXM2 GPUs connected through NVLink. The systems were organized into
four racks with 16 nodes each, connected via Infiniband. Each rack was equipped with a
TOR switch and racks were connected via a director switch. The systems used IBM Spectrum MPI
and PPC64LE Red Hat Enterprise Linux 7. As Deep Learning framework served IBM-Caffe (released
as part of PowerAI), which is derived from the open source bvlc-Caffe with enhancements customized 
for IBM's Power8 S822LC servers. Additionally, a PowerAI DDL enabled version of Torch was 
employed.

We performed two types of experiments - one to extract scaling efficiency as a function
of the number of GPUs, the other to report the run time to train a neural network
to attain desired accuracy. For the former experiment, the
batch size on each host was fixed, so that the total or effective batch size scaled with the number
of hosts. We chose an effective batch size that has been demonstrated to converge to best accuracy, so
the scaling efficiency reported is realistic.

We tested with two different networks, Resnet-50 for Imagenet with 1K image classes and
Resnet-101 \cite{RESNET} modified for Imagenet with 22K classes by increasing the width of the fully connected layer.

\subsection{Resnet-50 1K}

We used Resnet-50 to extract scaling efficiency and communication overheads for 4 GPUs, 8 GPUs, up to 256 GPUs,
while maintaining a fixed batch size of 32 per GPU. At 256 GPUs the effective batch size is 8192, the same
as Facebook AI Research's 256 GPU result \cite{FACEBOOK-256}. Since Facebook's paper has already demonstrated successful convergence to
best accuracy for 8192 batch size, the scaling efficiency number is meaningful.
The first Resnet-50 1K we used is a port of the network described in \cite{FACEBOOK-256} to Caffe.
Since Facebook did not publish the script for the Resnet-50 they used, we cannot guarantee that
our Resnet-50 is identical to theirs. However, both have about 100 Mbyte of gradients, so a
direct comparison on the communication overheads can be made. Since we used Caffe whereas Facebook
used Caffe2, giving rise to different run times per mini-batch iteration, the scaling efficiency cannot be directly compared.
The PowerAI DDL communication overhead of 64 Power8 S822LC servers (256 GPUs) 
over one Power8 S822LC server (4 GPUs) is about 23 ms.
Extracting from figure 7 of the Facebook paper, the communication overhead of 32 DGX-1 nodes (256 GPUs)
over 1 DGX-1 node (8 GPUs) is about 29 ms. Therefore our combined communication algorithm together with
hardware and network is slightly better.

\begin{figure}[h]
\begin{center}
\includegraphics*[width=5in]{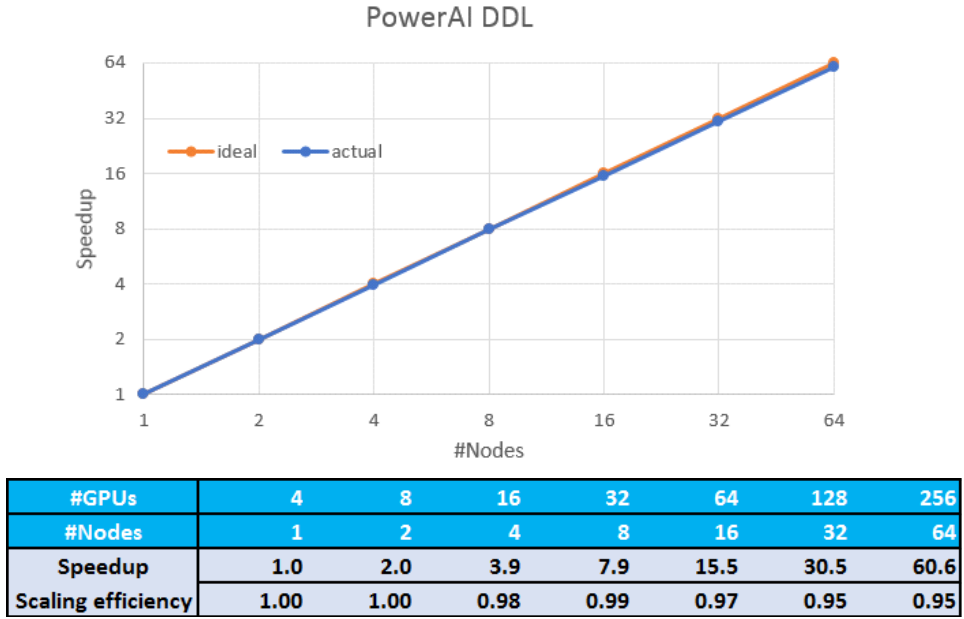}
\end{center}
\caption{Resnet-50 for 1K classes using up to 256 GPUs with Caffe.}
\label{fig-R192}
\end{figure}

Figure \ref{fig-R192} shows the speed up and scaling efficiency for different numbers of GPUs
for Resnet-50 with 1K image classes, with a batch size of 32 per GPU.
All runs in figure \ref{fig-R192} were performed with IBM-Caffe, and thus included
performance improvements to BVLC-Caffe. As a result, the intra-node communication between 4 MPI-ranks
within a single node performed better than the intra-node communication in multi-threaded native
BVLC-Caffe.

Furthermore, we trained Resnet-50 with 1K image classes with a PowerAI DDL enabled version of Torch \cite{torch} 
based on the open-source single-node multi-GPU Torch application available at \cite{soumith}.
A LUA interface to PowerAI DDL functionality for gradient reduction and distribution allows the adaption
with minimal modifications to the LUA implementation of Resnet-50 for 1K classes \cite{resnet}.
Learners are synchronized in every iteration. 

\begin{figure}[h]
\begin{center}
\includegraphics*[width=5in]{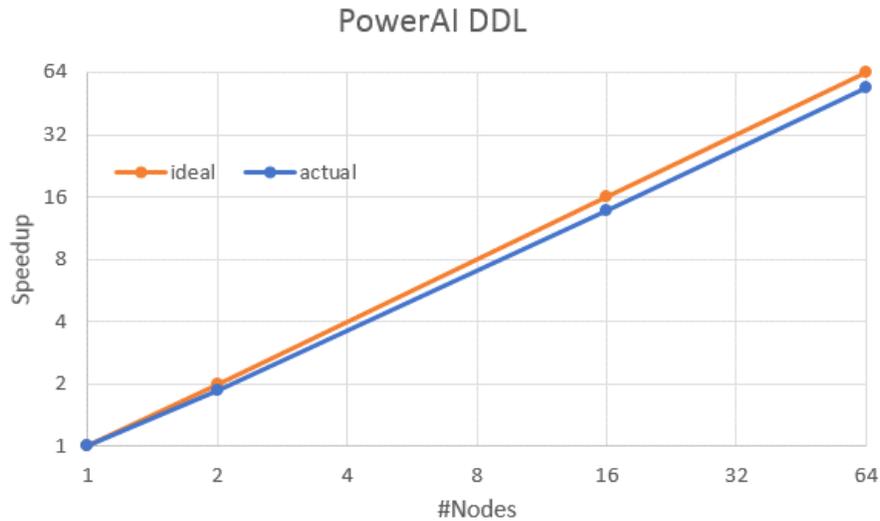}
\end{center}
\caption{Resnet-50 for 1K classes using up to 256 GPUs with Torch.}
\label{fig-R192-torch}
\end{figure}

Our implementation achieved an accuracy of 75.01\% in a training time of 50 minutes, training 
over 90 epochs with a batch size of 32 per GPU. The learning rates and weight-decay were chosen as in \cite{facebook}.
The original Resnet paper reported a peak accuracy of 75.3 \% on Resnet-50 \cite{resnet}. However, the results in \cite{facebook} show that the accuracy could be pushed to as high as 76.2 \% using hyper-parameter tuning for large effective batch-sizes. We have not incorporated those techniques in our work.

For these Torch Resnet-50 runs, figure \ref{fig-R192-torch} shows scalability across increasing numbers of GPUs. A scaling efficiency of 84\% is observed from 4 GPUs (single node) to 256 GPUs
(64 Nodes). It should also be noted that in the current Torch implementation, the reduction of gradients is not overlapped with compute.

\subsection{Resnet-101 22k}

\begin{figure}[h]
\begin{center}
\includegraphics*[width=5in]{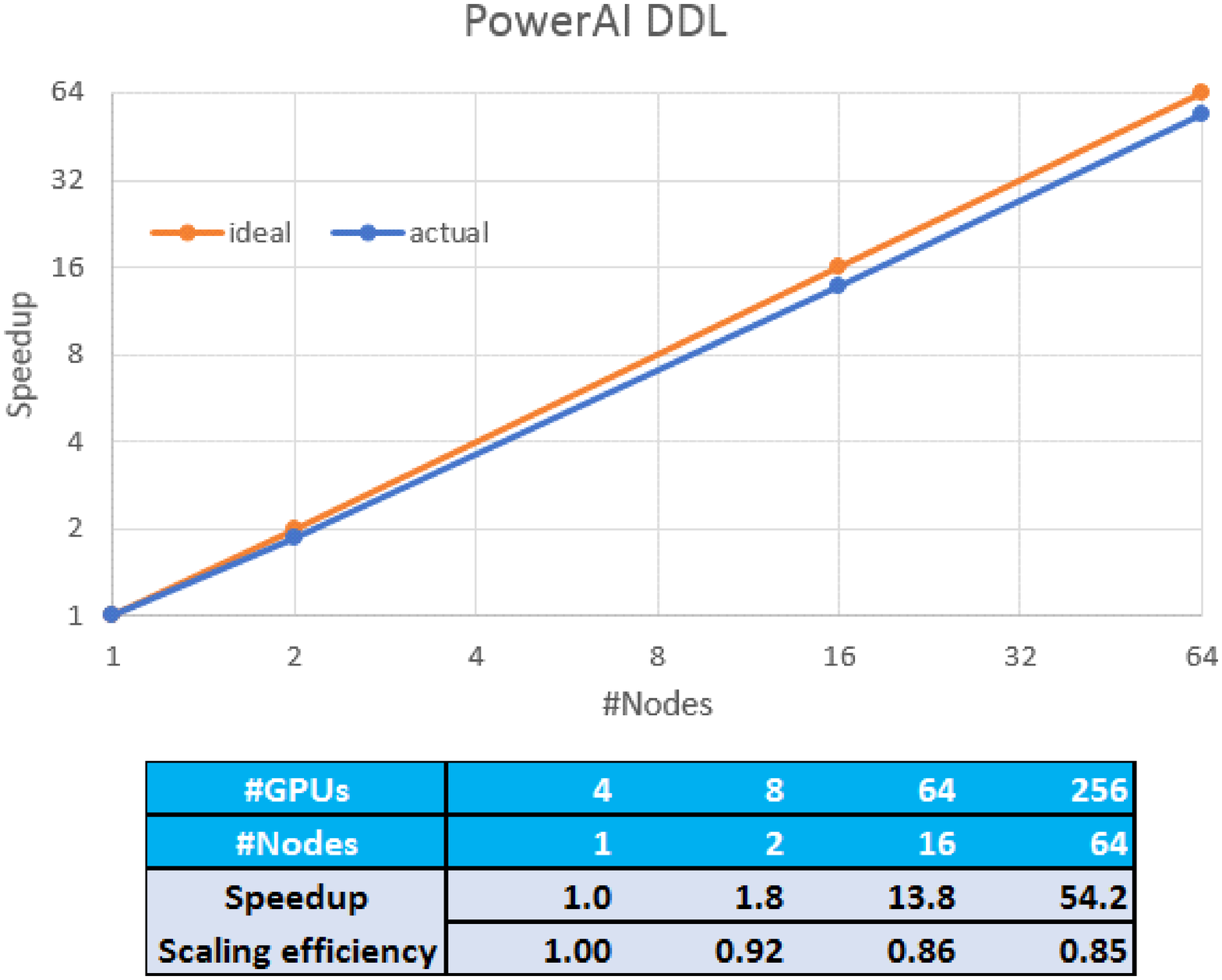}
\end{center}
\caption{Resnet-101 for 22k classes using up to 256 GPUs. }
\label{fig-R256}
\end{figure}

Compared to Resnet-50 with 1K image classes, 
Resnet-101 modified for 22K image classes has a roughly three times larger gradient set of 350 Mbyte, but the
corresponding increase in compute time for the forward/backward pass is not nearly as much.
Hence, it is a harder problem to scale due to the lower compute-to-communication ratio relative
to Resnet-50.
Figure \ref{fig-R256} shows our results running IBM-Caffe using 1, 2, 16, 32 and 64 Power8 S822LC servers with
a batch size of 20 per GPU. All runs in figure \ref{fig-R192} were performed with IBM-Caffe.

For batch size 5120 we achieved 33.8\% validation accuracy within about 7 hours. 
This accuracy exceeds the previous published accuracies of 29.8\%\cite{ADAM} and $<$25\%\cite{DistBelief} 
for Imagenet 22K. 
We used 7.5 million images from the full Imagenet dataset for training. Following standard practice, 
a separate dataset of 2 million images not used in training was used for validation.

\section{Conclusions} \label{sec-concl}

We have provided further confirmation that training complex deep neural networks with a large number of GPUs
using a synchronous data parallel methodology can scale almost linearly to hundreds of GPUs, thus
offering a viable strategy to reduce training time from weeks to hours. Such rapid turn around can
accelerate the improvement of existing neural networks and design of new neural networks, and
exploration of new application domains. To proliferate this technology to the masses, more research needs to be
done. Massive GPU scaling relies on successful training to good accuracy for large batch size. 
FAIR showed an initial demonstration of distribution to 256 GPUs on a couple of Resnets; our work here
demonstrates even better communication overhead, and uses PowerAI DDL to train a larger neural network
to leadership Imagenet 22K accuracy. It is key that the community continue to extend demonstration of
large-scale distributed Deep Learning to other popular neural network types, in particular, recurrent neural networks. The whole
training has to be made resilient and elastic since it is very likely that some devices will
malfunction when the number of learners increases. Automation and usability issues have to be addressed
to enable more turnkey operation, especially in a cloud environment.

\end{document}